\documentclass[12pt,letterpaper]{article}
 
\usepackage{amsfonts}
\usepackage{graphicx}
\usepackage{amsmath}

\def\be{\begin{equation}}
\def\ee{\end{equation}}
\def\bear{\begin{eqnarray}}
\def\eear{\end{eqnarray}}
\def\nn{\nonumber}

\def\half{{{1\over 2}}}

\newcommand\inv[1]{{1\over{#1}}}

\def \a{\alpha}
\def \s{\sigma}
\def \ds{\partial_\sigma}
\def \Tr{\mathrm{Tr}}
\begin{document}

%%%%%%%%%%%%%%%%%%%%%%%%%%%%%%%%%%%%%%%%%%%%%%%%%%%%%%%%%%%%%%%%%%%%
%  TITLE PAGE                                                      %
%%%%%%%%%%%%%%%%%%%%%%%%%%%%%%%%%%%%%%%%%%%%%%%%%%%%%%%%%%%%%%%%%%%%
%  Find all ***                                                    %
%%%%%%%%%%%%%%%%%%%%%%%%%%%%%%%%%%%%%%%%%%%%%%%%%%%%%%%%%%%%%%%%%%%%

\begin{titlepage}
~\vskip 1in
\begin{center}
{\Large
{Noncommutative Geometry of Multicore Bions}}
\vskip 0.5in
{Joanna L. Karczmarek and Ariel Sibilia}
\vskip 0.3in
{\it 
Department of Physics and Astronomy\\
University of British Columbia\\
Vancouver, Canada}
\end{center}

\vskip 0.5in
\begin{abstract}
We find new BPS solutions to the nonabelian theory on a world-volume
of parallel D1-branes.  Our solutions describe two parallel,
separated bundles of $N$ D1-branes expanding out to form a single orthogonal 
D3-brane.  This configuration corresponds to two charge $N$ magnetic monopoles
in the world-volume of a single D3-brane,
deforming the D3-brane into two parallel spikes.
We obtain the emergent surface corresponding to our nonabelian
D1-brane configuration and demonstrate, at finite $N$,
a surprisingly accurate agreement with the shape of the D3-brane world-volume 
as obtained from the abelian Born-Infeld action.  Our solution
provides an explicit realization of topology change in noncommutative
geometry at finite $N$.
\end{abstract}
\end{titlepage}

\tableofcontents

%%%%%%%%%%%%%%%%%%%%%%%%%%%%%%%%%%%%%%%%%%%%%%%%%%%%%%%%%%%%%%%%%%%%
%  DRAFTMORE ONLY; ADD MORE AUTHORS IF NEEDED                      %
%%%%%%%%%%%%%%%%%%%%%%%%%%%%%%%%%%%%%%%%%%%%%%%%%%%%%%%%%%%%%%%%%%%%

\section{Introduction and Summary}
\label{sec:intro}

Since the discovery that the coordinate positions of D-branes
in string theory are matrices, many aspects of D-brane dynamics
related to this fact have been studied. The gauge group
of a theory living on a stack of $N$  superposed D-branes is enhanced
from $U(1)^N$ to $U(N)$ and  the brane world-volume supports a $U(N)$
gauge field as well as a set of scalars in the adjoint representation
of $U(N)$ (one for each of the transverse coordinates).
A striking feature of the nonabelian Born-Infeld effective action
for stacks of D-branes \cite{Tseytlin:1997cs,Myers:1999ps} is that it contains
structures such as commutators of the transverse coordinates
with themselves. These vanish in the $U(1)$ case and cannot be directly
inferred from the abelian Born-Infeld action.
They lead to nonabelian geometrical structures and allow
lower dimensional branes to carry higher dimensional brane charges,
so that higher dimensional branes can be built from lower dimensional branes.

Thus,  higher dimensional objects can be built from
lower dimensional components.  This approach complements the opposite view that
lower dimensional branes can be described as solitons in the effective theory 
of  higher dimensional D-branes \cite{Douglas:1995bn}.
Where higher dimensional D-branes emerge from lower dimensional ones,
properties of nonabelian configurations of
lower dimensional branes can be compared against the properties of the
corresponding higher dimensional branes.

\begin{figure}
\centering
\includegraphics[width=0.4\textwidth]{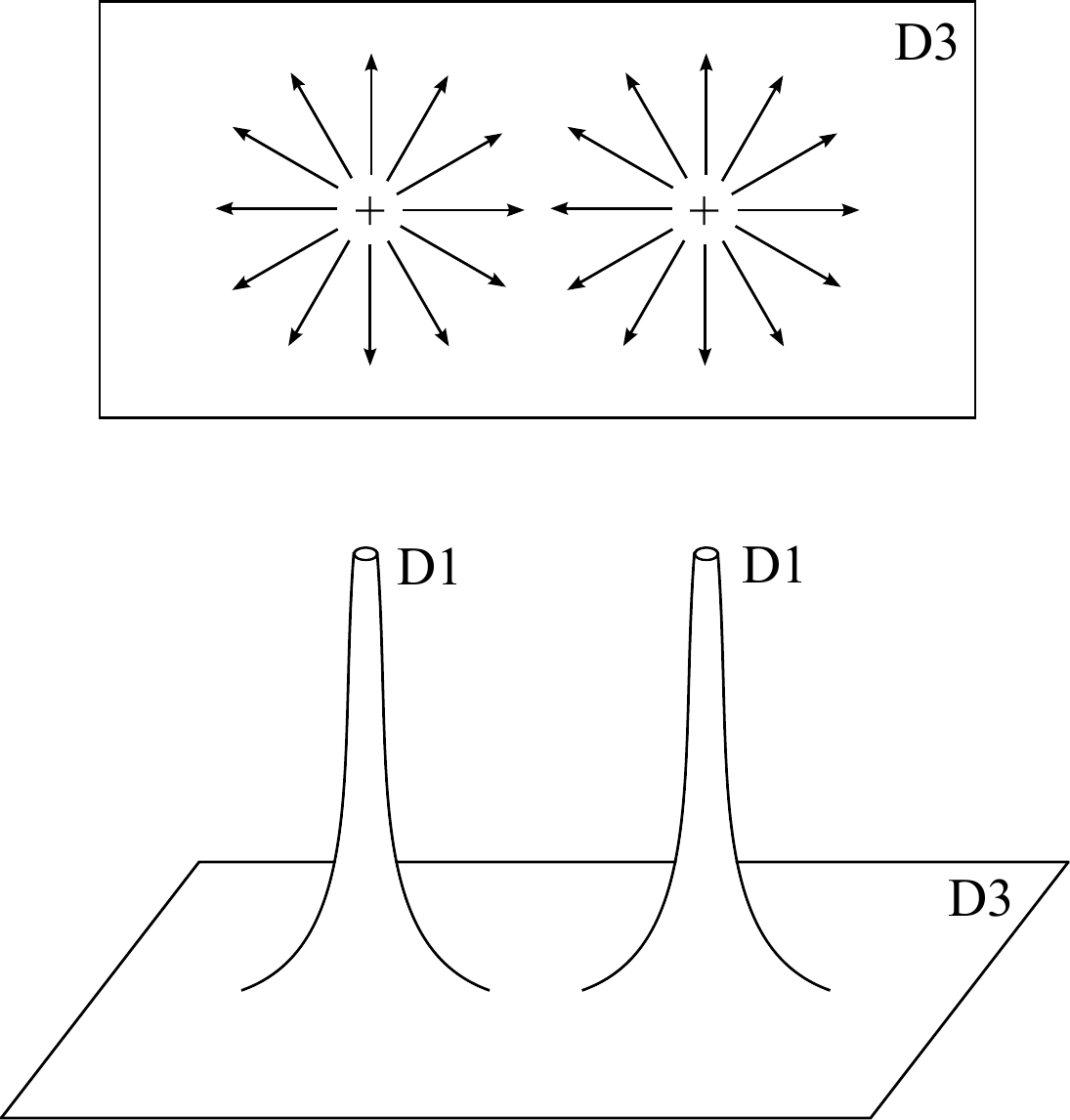}
\caption{A D3-brane with two magnetic monopoles.  The nonlinearity of the 
Born-Infeld action deforms the D3-brane so that spikes form at the locations of 
the monopoles.  These correspond to bundles of attached D1 branes.}
\label{fig:spikes}
\end{figure}

One well studied example is the D1-D3 brane intersection.  This can be described from
either the D3 \cite{Callan:1997kz,Gibbons:1997xz}
 or the D1 brane point of view \cite{Constable:1999ac}.  From the D3 brane point of view,
the point where the D1 branes are attached is a magnetic monopole (see Figure
\ref{fig:spikes}).  A single D3 brane in flat space can be described by the abelian Born-Infeld action. 
Let the D3-brane extend in 0123-directions,
and let the coordinates on the brane be denoted by $x^i$,
$i=0,\ldots,3$. Restricting the brane to have displacement only in
one of the transverse directions, we can take the (static gauge)
embedding coordinates of the brane in the ten-dimensional space to be
$X^i = x^i$, $i=0,\ldots,3$; $X^a = 0$, $a=4,\ldots,8$;
$X^9 = \sigma(x^i)$. Then there exists static solutions of
the Born-Infeld action with $k$ `spikes', corresponding to placing  $N_{(a)}$ units of
$U(1)$ magnetic charge at positions $x^i_{(a)}$, for $a$ from $1$
to $k$ \cite{Callan:1997kz}:
\be
X^9(x^i) = \sigma(x^i) =
\sum_{a=1}^{k}~
{q_{(a)} \over {\sqrt{(x^1-x^1_{(a)})^2+(x^2-x^2_{(a)})^2+(x^3-x^3_{(a)})^2}}} ~,
\label{eqn:bions}
\ee
where $q_{(a)}=\pi\alpha' N_{(a)}$, and $N_{(a)}$ are integers corresponding
to the number of D1-branes attached at point $x^i_{(a)}$. We have
omitted the world-volume magnetic field, as it will not enter into our discussion.
This solution is reliable
in the sense that the effect of unknown higher-order corrections in
$\alpha'$ and $g$ to the action can be made systematically small
in the large-$N$ limit (see \cite{Callan:1997kz} for details).
From the point of view of the D3-brane, the magnetic monopoles are
BPS objects and thus there is no net force between them when
placed at a finite distance from each other.  Thus, the family 
of solutions in (\ref{eqn:bions}) has a very large moduli space:
an arbitrary superposition of `spikes' shown in equation (\ref{eqn:bions}) is a solution.

The same D1-D3 intersection can be described from the point of view of D1-strings.
The goal of this paper is to study the properties of such a description for
multiple separated D1-string bundles (multiple D3-brane spikes). We will use static gauge,  
and have the D1-branes stretching along the $X^9$ direction,
with $X^0=\tau, ~X^9 = \sigma$.  We allow three of the transverse coordinates
to be nonzero.  These are described by  matrix transverse
scalar fields $\Phi^i(\sigma)$, for $i=1,2,3$.  We will see in Section \ref{sec:nahm}
that $\Phi^i(\sigma)$ corresponding to multiple spikes attached to a single 
D3-brane can be obtained by solving the Nahm equation with
novel boundary conditions.  Once that is accomplished, we need to compare the emergent
geometry in the D1-brane picture with the geometry of the D3-brane.  
To this end, consider a cross-section of the D3-brane 
at fixed $X^9$.  This is a codimension one surface in the three dimensional
subspace given by $X^1$, $X^2$ and $X^3$.  This surface will correspond to
the nonabelian object described by the three coordinate matrices $\Phi^i(\sigma)$.

Before describing our topologically and metrically nontrivial nonabelian geometry, consider
the case where this surface is simply a round sphere.  This is the 
well-studied example of a single bundle of $N$ D1-branes \cite{Constable:1999ac}.
The cross-section of the D3-brane 
at fixed $X^9$ is a 2-sphere whose radius varies with $\sigma=X^9$ as $\pi\alpha' N/\sigma$.
In the D1-brane picture, the solution is (see Section \ref{sec:nahm} for details)
\be
\Phi^i = \frac{1}{\sigma} \alpha^i~,
\label{eqn:one-bundle-1}
\ee
where $\alpha^i$ are the $N\times N$-dimensional
generators of an irreducible representation of $SU(2)$, with
$[\alpha^i, \alpha^j] = i\epsilon_{ijk}\alpha^k$.  One way to compute the 
radius of this noncommutative sphere is from the quadratic Casimir:
$(R(\sigma))^2 = {(2\pi\a')^2\over N}\Tr (\Phi^i)^2$, which
leads to $R(\sigma) = \pi\alpha' \sqrt {N^2-1}/\sigma$, in agreement
with the D3-brane picture at large $N$.

\begin{figure}
\centering
\includegraphics[width=0.70\textwidth]{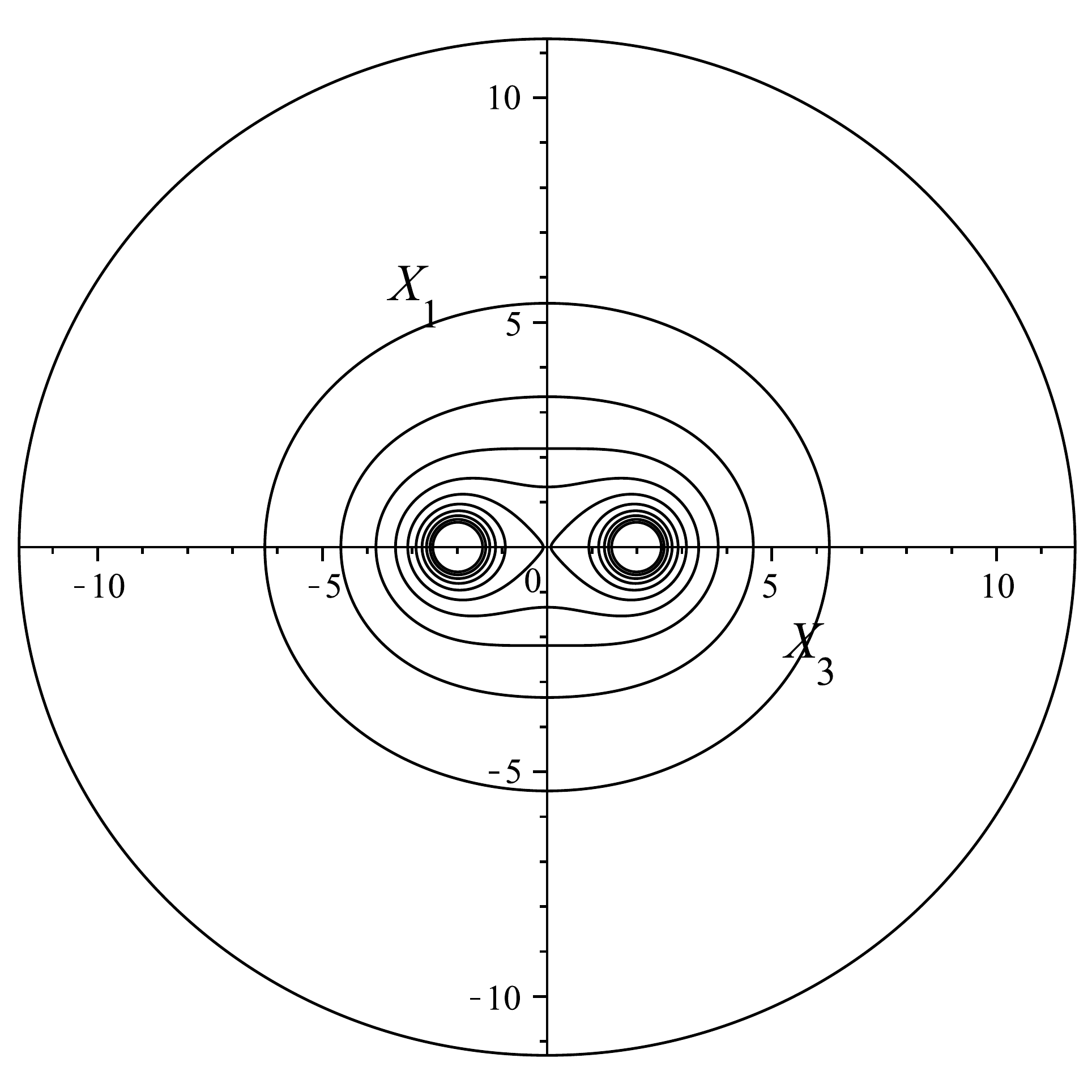}
\caption{Surfaces emerging from the nonabelian configuration of D1-branes
at different points along the D1-brane world-volume, parametrized by $\sigma$.
The outer surface correspond to smallest
$\sigma$, while the most-inner two circles correspond to largest $\sigma$, with
$\sigma$  increasing linearly with surface number.  The topological transition from a single
2-sphere to two 2-spheres is clearly demonstrated. 
We should stress that these smooth surfaces are obtained solely from the matrix
transverse coordinates of the D1-brane.
($X^2=0$ cross-sections of co-dimension one surfaces are shown.
The nonabelian configuration of D1-brane bundles was found numerically (see Section \ref{sec:soln}).  
There are two D1-brane bundles separated in the $X^3$ direction, each containing 6 D1-branes.)}
\label{fig:contours}
\end{figure}

For multiple D1-brane bundles, the cross-section at fixed $X^9$ is more
complicated, as are the  corresponding D1-string configurations, $\Phi^i(\sigma)$.
In Section \ref{sec:soln}, we will describe a method for numerically computing $\Phi^i(\sigma)$.  
Once such a nonabelian configuration is obtained, a method is needed to reconstruct the
corresponding geometry.  Fortunately, a tool to do just that has
recently been developed \cite{Berenstein:2012ts}: given a set of three
hermitian $N\times N$ matrices, the corresponding geometry in 
the flat three dimensional space parametrized by $(X^1, X^2, X^3)$ is given
by the locus of points where a certain effective Hamiltonian
has a zero eigenvalue. The relevant effective Hamiltonian acts on
a $2N$-dimensional auxiliary space and is given by
\be
H_{\mathrm{eff}} = \sum_{i=1}^3 \left (\Phi^i - X^i \right ) \otimes \sigma_i~,
\label{eqn:Heff}
\ee
where $\sigma_i$ are the $2\times 2$ Pauli matrices.  Using this method,
we are able to find the emergent geometry corresponding to our D1-brane solutions.
An example, with two D1-brane bundles, is shown in Figure \ref{fig:contours}.  
The contours seen in the Figure were obtained solely from
the nonabelian data describing D1-branes.  They have all the qualitative
features expected from a contour plot of equation (\ref{eqn:bions}) for $k=2$.  
A detailed comparison between the prediction of equation (\ref{eqn:bions}) and the
emergent surfaces  is presented in Figures \ref{fig:contours-compare}, \ref{fig:profile}, and
\ref{fig:saddle-point}.

The results presented here are interesting for several reasons.  We demonstrate that the
correspondence between the D3- and the D1-brane description of the D1-D3 brane intersection
is applicable beyond the simple example of a single D1-brane bundle.
We move beyond solutions based on round noncommutative spheres and 
explicitly solve for a nontrivial nonabelian geometry arising on a bundle of D1-branes,
demonstrating that our solution smoothly interpolates between a block diagonal configuration
corresponding to two separated bundles and an irreducible configuration corresponding to the 
D1-branes flaring out to form a single D3-brane.  This smooth interpolation constitutes an example
of a nonabelian topology change, with the `universe' changing its topology from a single 
2-sphere to two disjointed 2-spheres, clearly seen in Figure \ref{fig:contours}.

A curious feature of the correspondence between the D3- and the D1-brane picture
is that, since the BPS condition for a D3-brane is linear, in the D3-brane
picture the `spikes' (or D1-brane bundles) are non-interacting and a
solution involving any number of them is simply a superposition of the solutions for the individual
`spikes'.  In contrast, the BPS condition for the D1-branes (the Nahm equation) is
nonlinear and it is not at all a priori clear that when several D1-brane bundles flare out to
form one D3-brane they remain non-interacting.  The existence of our solutions
shows that this is indeed the case; it would be interesting to see in detail how 
the large moduli space of  BPS states on a D3-brane arises from the non-linear Nahm equation.

\section{D1-branes and the Nahm equation}
\label{sec:nahm}

In this section, we will describe the D1-D3 brane intersection in terms of
the D1-brane action \cite{Constable:1999ac}.

We consider the nonabelian Born--Infeld action 
specialized to the case of $N$ coincident D1-branes,
in a flat background spacetime, with vanishing
$B$ field, vanishing world-volume gauge field and constant dilaton.
The action depends only on the  matrix transverse
scalar fields $\Phi^i$'s. In general, $i=1,\ldots,8$, but since we are
interested in studying the D1/D3-brane intersection, we will allow
only three transverse coordinate fields to be active ($i=1,2,3$).
The explicit reduction of the static gauge action ($X^0 = \tau$ and
$X^9=\sigma$) is then \cite{Tseytlin:1997cs}
\be
\label{eqn:SBI1}
S_{BI} = -T_1 \int d\sigma d\tau \mathrm{STr}
 \sqrt{ -\det(\eta_{ab}+(2\pi\alpha')^2\partial_a\Phi^i
   Q^{-1}_{ij}\partial_b\Phi^j) det(Q^{ij}) } ~,
\ee
where
\be
Q^{ij} = \delta^{ij} + i 2\pi\a' [\Phi^i,\Phi^j]~.
\ee
Since the dilaton is constant, we incorporate it in the
tension $T_1$ as a factor of $g^{-1}$.

The solution we are interested in is a static BPS solution.
It can be argued \cite{Constable:1999ac} that the BPS condition is
the Nahm equation,
\be
\partial \Phi_i =  \half i \epsilon_{ijk}[\Phi^j,\Phi^k] ~.
\label{eqn:BPS}
\ee

The trivial solution has  $\Phi^{ij}=0$, which
corresponds to an infinitely long bundle of coincident D1-branes.
In \cite{Constable:1999ac}, a more interesting solution was found,
starting with the following ansatz:
\be
\Phi^i = 2 \hat R(\sigma) \alpha^i,
\label{eqn:ansatz0}
\ee
where $\alpha^i$ are the generators of some representation of the $SU(2)$ subgroup of $U(N)$,
$[\alpha^i, \alpha^j] = i\epsilon_{ijk}\alpha^k$.
When this ansatz is substituted into the BPS condition
(\ref{eqn:BPS}), we obtain a simple equation for $\hat R$,
\be
\hat R'= - 2\hat R^2 ~,
\ee
which is solved by
\be
\hat R =  \inv{2 \sigma}~.
\label{eqn:R}
\ee

As was already discussed in the Introduction, for an irreducible
representation of $SU(2)$ with dimension $N$, this
solution maps very nicely onto the single bion solution of the D3-brane action.
At a fixed point $|\sigma|$ on the D1-brane stack,
the geometry given by (\ref{eqn:ansatz0}) is that of a sphere
with the physical radius at large $N$ equal to
$R(\sigma)= {{\pi\a' N } / {(|\sigma|)}}$.

We now would like to obtain an analog of the
multi-bion solutions (\ref{eqn:bions}) in the D1-brane description.
This requires finding a solution to the BPS equation
(\ref{eqn:BPS}) with the following properties
\footnote{We are proposing these boundary conditions based
on geometrical considerations alone, but it should be possible
to derive them in a way similar to \cite{Chen:2002vb}.}. 
At $\s \rightarrow
\infty$, we would like a number of parallel D1-brane bunches,
separated in space, therefore the three matrices $\Phi^i$ should be
of the block diagonal form: \be \label{eqn:spikes} \Phi^i(\sigma \rightarrow \infty)
\sim {\mbox{diag}} \left( x^i_{(1)} {\bf I}_{q_{(1)}\times q_{(1)}}
+ 2 \hat R ~\alpha^i_{q_{(1)}}, ~~\dots,~~ x^i_{(k)} {\bf
I}_{q_{(k)}\times q_{(k)}} + 2 \hat R ~\alpha^i_{q_{(k)}} \right )~,
 \ee where $\alpha^i_{q_{(a)}}$ are
$q_{(a)}$-dimensional generators of an irreducible representation of
$SU(2)$. At $\s=0$, though, the residue must be an  {\it irreducible}
representation of $SU(2)$, so that a {\it single} D3-brane is formed:
\be \label{eqn:brane} \Phi^i(\sigma \rightarrow 0) \sim 2 \hat R
~\alpha^i_{q_{(1)}+\ldots +q_{(k)}} ~+~{\textrm{finite}},  \ee

The desired solution  to equation (\ref{eqn:BPS}) must then
interpolate between $k$ irreducible representations
of $SU(2)$ at large $\s$, and a single irreducible representation
at small $\s$.  This corresponds to topology change in the 
emergent noncommutative geometry.

The expected correspondence between the two descriptions of 
the BPS D1-D3 brane intersection
strongly suggests that such solutions exists for arbitrary
$k$, $q_{(a)}$s and $x^i_{(a)}$s.  In the next section we will
explicitly solve a simple but nontrivial example with $k$=2 and
$q_{(1)}=q_{(2)}$.

\section{Two separated D1-brane bundles}
\label{sec:soln}

Consider two D1-brane $N$ bundles, parallel but separated by a distance $D$
in the $X^3$ direction.  The solution to the Nahm equation (\ref{eqn:BPS})
we seek must have the behaviour
\be
\Phi^i(\sigma \rightarrow \infty) \sim 
\left [\begin{array}{cc}
\dfrac{\alpha^i_N}{\s} + \delta_{3,i} \dfrac{D}{2} & 0 \\ 
0 & \dfrac{\alpha^i_N}{\s}  -\delta_{3,i} \dfrac{D}{2} 
\end{array} \right ]
\label{eqn:at-infty}
\ee
and 
\be
\Phi^i(\sigma \rightarrow 0) \sim  \dfrac{\alpha^i_{2N}}{\s}~.
\label{eqn:at-zero}
\ee
We use the following ansatz 
\be
\Phi^1 = \left [\begin{matrix}
0 & A_{1}  & ~ & ~ & ~ & ~ \\
A_1 & \ddots  & \ddots & ~ & ~ & ~\\
~ & \ddots & ~& A_{N} & ~ & ~ \\
~ &~ & A_N  & ~ & \ddots & ~  \\
~ & ~ & ~ &  \ddots & \ddots  & A_{1}\\
 ~ & ~ & ~ & ~ & A_1 & 0   \\
\end{matrix} \right ]~,
\qquad
\Phi^2 = \left [\begin{matrix}
0 & iA_{1}  & ~ & ~ & ~ & ~ \\
-iA_1 & \ddots  & \ddots & ~ & ~ & ~\\
~ & \ddots & ~& iA_{N} & ~ & ~ \\
~ &~ & -iA_N  & ~ & \ddots & ~  \\
~ & ~ & ~ &  \ddots & \ddots  & iA_{1}\\
 ~ & ~ & ~ & ~ & -iA_1 & 0   \\
\end{matrix} \right ] ~,
\nn \ee
\be 
\Phi^3 = \left [\begin{matrix}
A_{N+1} & ~ & ~ & ~ & ~ & ~ \\
~ & \ddots & ~ & ~ & ~& ~\\
~ & ~ & A_{2N} & ~ & ~ & ~\\
~ &~ & ~ & A_{2N} & ~ & ~ \\
~ & ~ & ~ &  ~ &\ddots  & ~\\
 ~ & ~ & ~ & ~ & ~ &  A_{N+1}  \\
\end{matrix} \right ]~.
\ee
We are faced with the problem of numerically solving a set of differential equation
in the case where some boundary conditions are specified at $\sigma=0$ and some
are specified at $\sigma=\infty$, with the added complication that 
the equations are singular at both  $\sigma=0$  and $\sigma=\infty$.  Our
solution to this problem is to begin the numerical  integration a small but 
nonzero value of $\sigma$.  To obtain the initial conditions of the numerical
solution, we use a series expansion at $\sigma=0$.  We ensure that the coefficients of 
this expansion `know' about the boundary condition at $\sigma=\infty$,
by enforcing conservation of certain quantities which we compute at
$\sigma=\infty$ and which depend on the separation of the 
two D1-brane bundles.

\begin{figure}
\centering
\includegraphics[width=0.6\textwidth]{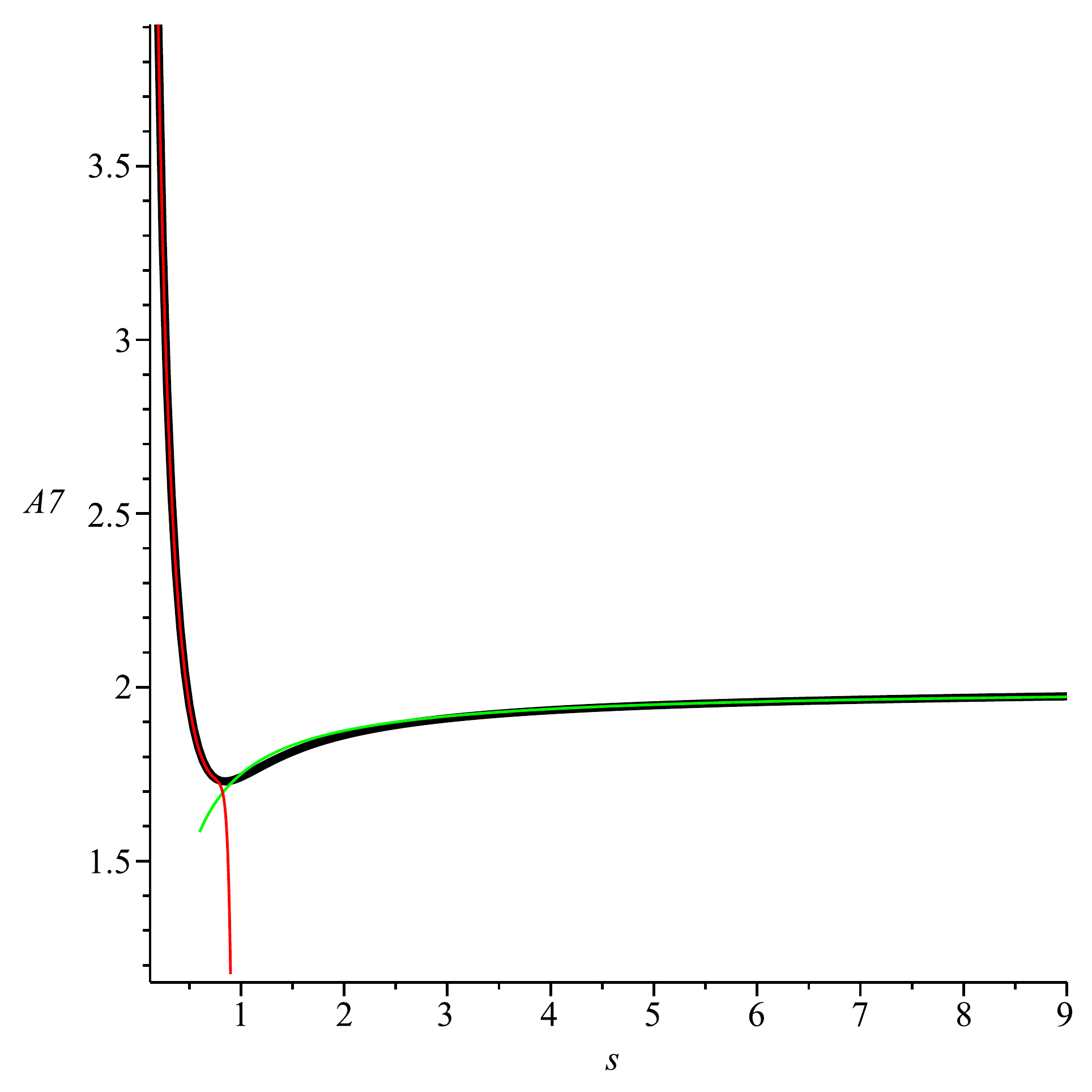}
\caption{$A_7$ as a function of $\s$, 
for $N=4$ and $D=4$.  Black: numerical solution, red: the 
small $\s$ expansion (\ref{eqn:expansion}) to order $\s^{31}$,
green: the asymptotic behaviour at large $\sigma$, 
equation (\ref{eqn:at-infty}).}
\label{fig:example}
\end{figure}

It is easy to see that we want the $2N$ undetermined functions 
$A_p$ to be odd in $\s$. Using the boundary condition (\ref{eqn:at-zero}) at small 
$\sigma$,
we fix the leading coefficients $c_{p,-1}$ in the following expansion
\be
A_{p} = \frac{c_{p,-1}}{\s} + c_{p,1}\s^1 + c_{p,3}\s^3 + c_{p,5}\s^5 + 
\ldots~.
\label{eqn:expansion}
\ee
Then, from the Nahm equation, all but $N$ of coefficients $c_{p,j}$, 
$j \ge 1$ can be obtained recursively.  The remaining $N$ coefficients (which
can be conveniently chosen to be, $c_{p,j}$, $j=1 \ldots 2N - 1$),
represent freedom remaining in boundary conditions at $\sigma = 0$ after
condition (\ref{eqn:at-zero}) is enforced.  This remaining freedom allows us
to satisfy our other boundary condition (\ref{eqn:at-infty}) at large $\sigma$.
To implement those, we take advantage of the well known fact that 
the Nahm equation can be rewritten in Lax form\footnote
{This is a special case, sufficient for out purpose;
for a more general case,
see for example a review in \cite{Sutcliffe:1997ec}.}
\be
M \equiv \half \left ( \Phi^1 + i \Phi^2 \right )
~,~~ L \equiv  \half \Phi^3~,~~
\ds M =  [M ,L]~.
\label{eqn:lax}
\ee
$\Tr (M^k)$ is a constant for any k.  By substituting
the asymptotic behaviour (\ref{eqn:at-infty}) into $\Tr (M^k)$ for
$k = 2, 4, \ldots, 2N$ we obtain $N$ constants of motion which allow us
to solve for the remaining coefficients in equation (\ref{eqn:expansion}).
The entire computation can be carried out efficiently in a symbolic
manipulation program such as Maple, and all coefficients can be 
obtained exactly.  Once we have  obtained expansion  (\ref{eqn:expansion}) to a sufficiently
high order, we can numerically integrate the Nahm equation for $\Phi^i(\s)$
from a starting point given by the expansion at some $\sigma_0$ near zero.
Figure \ref{fig:example}
shows a sample result from this procedure.  It is clear that our
solution correctly interpolates between the desired  small and large $\s$ 
asymptotic behaviour.  We should notice, however, that at larger $\s$ values
numerical integration leads to a divergence, and that the  region of its apparent validity 
depends on the starting point $\s_0$.  The larger $\s_0$, the larger the region of validity,
therefore, care must be taken to use a sufficiently high expansion order.

\section{Comparison with D3-brane picture}
\label{sec:geometry}
 
As described in the Introduction, we  use the recent technique of 
\cite{Berenstein:2012ts} to obtain the geometry corresponding to any
three hermitian matrices, and thus to compute the smooth D3-brane surface
that emerges from our D1-brane solutions.  We should note that while
work \cite{Berenstein:2012ts} derives equation (\ref{eqn:Heff}) for
nonabelian configurations of D0-branes, equation (\ref{eqn:Heff}) itself
should have more general applicability, as it simply provides a 
geometric interpretation in ${\mathbb R}^3$ of a set of three matrices.
We therefore apply it to the transverse coordinates of D1-branes
at a fixed point along their length.  The locus of points where $H_{\mathrm{eff}}$
vanishes corresponds then to a constant $\sigma$ contour of the emergent
D3-brane.  The shape of such contours is shown in Figure \ref{fig:contours}.

To obtain a meaningful, quantitative comparison with equation (\ref{eqn:bions}), it will
be useful first to review how the round noncommutative sphere arises from equation
(\ref{eqn:Heff}).
Consider a `unit' noncommutative sphere given by
\be
\Phi^i = \alpha^i/j~,
\ee
where $j=(N-1)/2$ is the spin of the $N$-dimensional irrep of $SU(2)$.
The Casimir of such an irrep is $\Tr(\Phi^i)^2 = (N^2-1)/4j^2 = (N+1)/(N-1)$,
previously leading us to believe that the corresponding sphere has radius
$\pi \a' \sqrt {N+1\over N-1}$.
However, when one considers the locus of points where $H_{\mathrm{eff}}$
vanishes, one gets a sphere of radius exactly $1$ \cite{Berenstein:2012ts}.
This implies that the solution (\ref{eqn:one-bundle-1}) (or
(\ref{eqn:ansatz0})), when viewed through the methodology of
\cite{Berenstein:2012ts}, corresponds to a sphere of radius $2 \pi\a' j /\s =
\pi\a' (N-1) /\s$.  To obtain a more meaningful comparison with equation
(\ref{eqn:bions}), we will thus use a modified charge: $q=\pi\a' (N-1)$ instead
of $q=\pi\a' N$.  This will allow us to obtain a comparison between our solution
and the imposed boundary conditions that is not obscured by the finite $N$ 
limitations of equation (\ref{eqn:Heff}).

\begin{figure}
\centering
{\bf (a)} \includegraphics[width=0.6\textwidth]{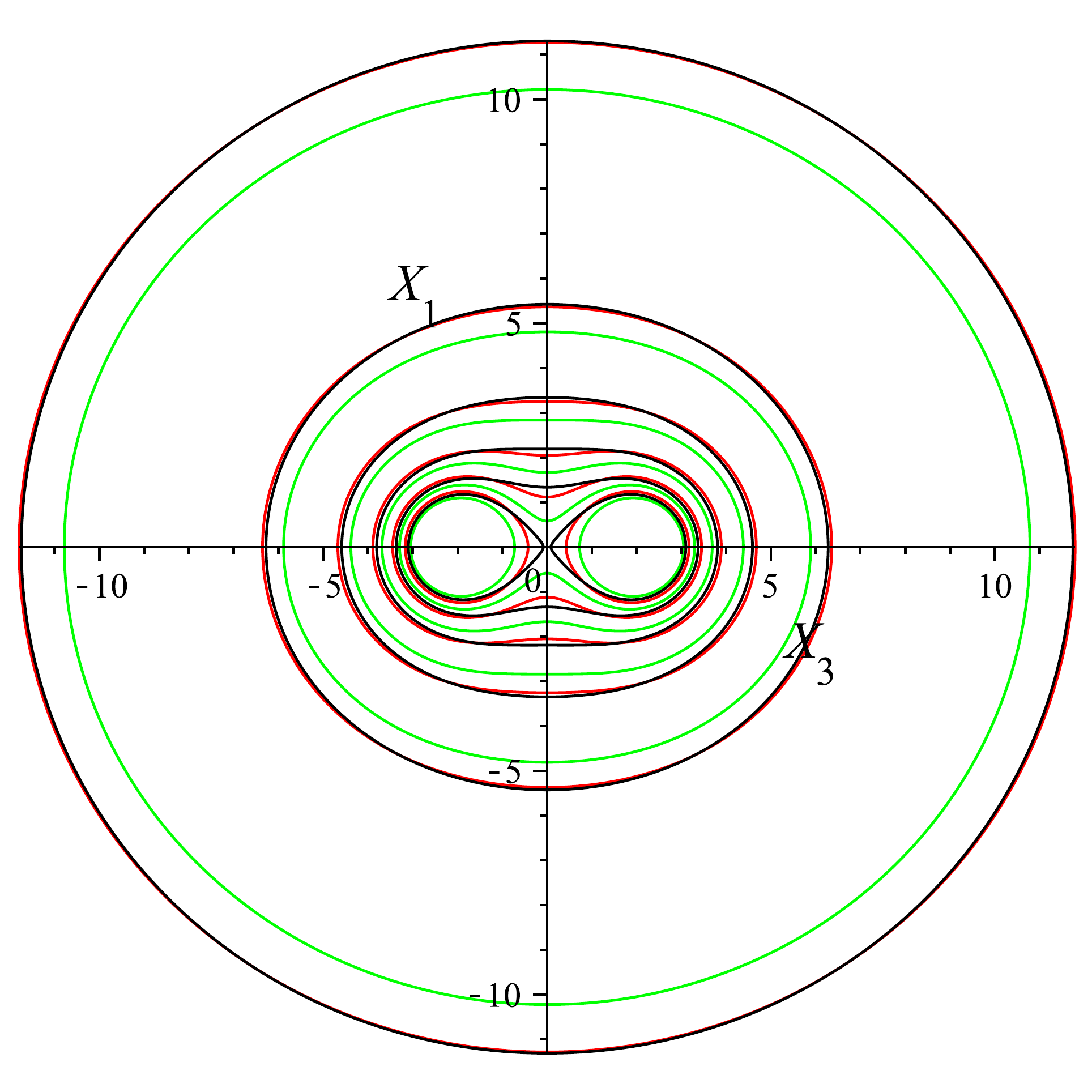}
\\
{\bf(b)} \includegraphics[width=0.6\textwidth]{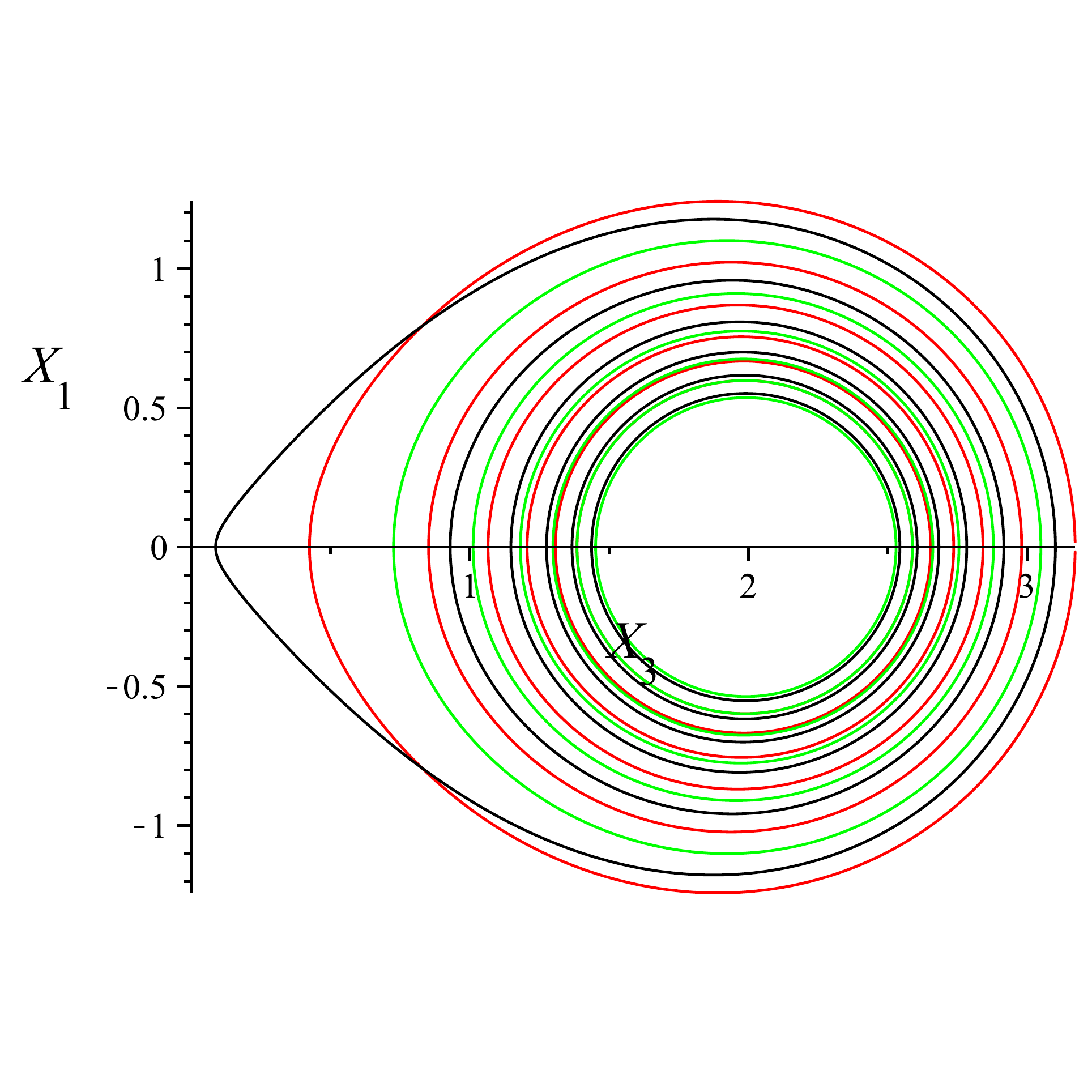}
\caption{Cross-sections as in Figure \ref{fig:contours}.  Black lines
from the D1 brane picture, red lines from the D3 brane picture, equation
(\ref{eqn:bions2}), with $q=N-1/2$ and green lines from the D3 brane picture,
equation (\ref{eqn:bions2}), with $q=N-1$. $N=6$.}  
\label{fig:contours-compare}
\end{figure}

The basic comparison is shown in Figure \ref{fig:contours-compare}.
Black contours are those previously shown in Figure \ref{fig:contours}.
The red and green lines are level sets of equation (\ref{eqn:bions}),
which for $k=2$ and $q_{(1)} = q_{(2)}= q$ reads simply
\be
\s(x^i) = \frac{q}{\sqrt{(x^1)^2+ (x^2)^2 +(x^3-D/2)^2 }} +
\frac{q}{\sqrt{(x^1)^2+ (x^2)^2 +(x^3+D/2)^2 }}~.
\label{eqn:bions2}
\ee
The green lines are plotted using $q=N-1$, which is appropriate
when each of the two bundles of $N$ D1-strings forms its own sphere.
The red lines are plotted using $q = (2N-1)/2$, which corresponds to
half the modified charge of a single $2N$ D1-string bundle.  These should
be more accurate away from the monopole (at large radius), where all $2N$ D1-branes
form a single sphere.  We see in Figure \ref{fig:contours-compare} that the
agreement is excellent: at large radius the black lines coincide with the red lines
while at small radius the black lines coincide with the green lines.

\begin{figure}
\centering
\includegraphics[width=0.5\textwidth]{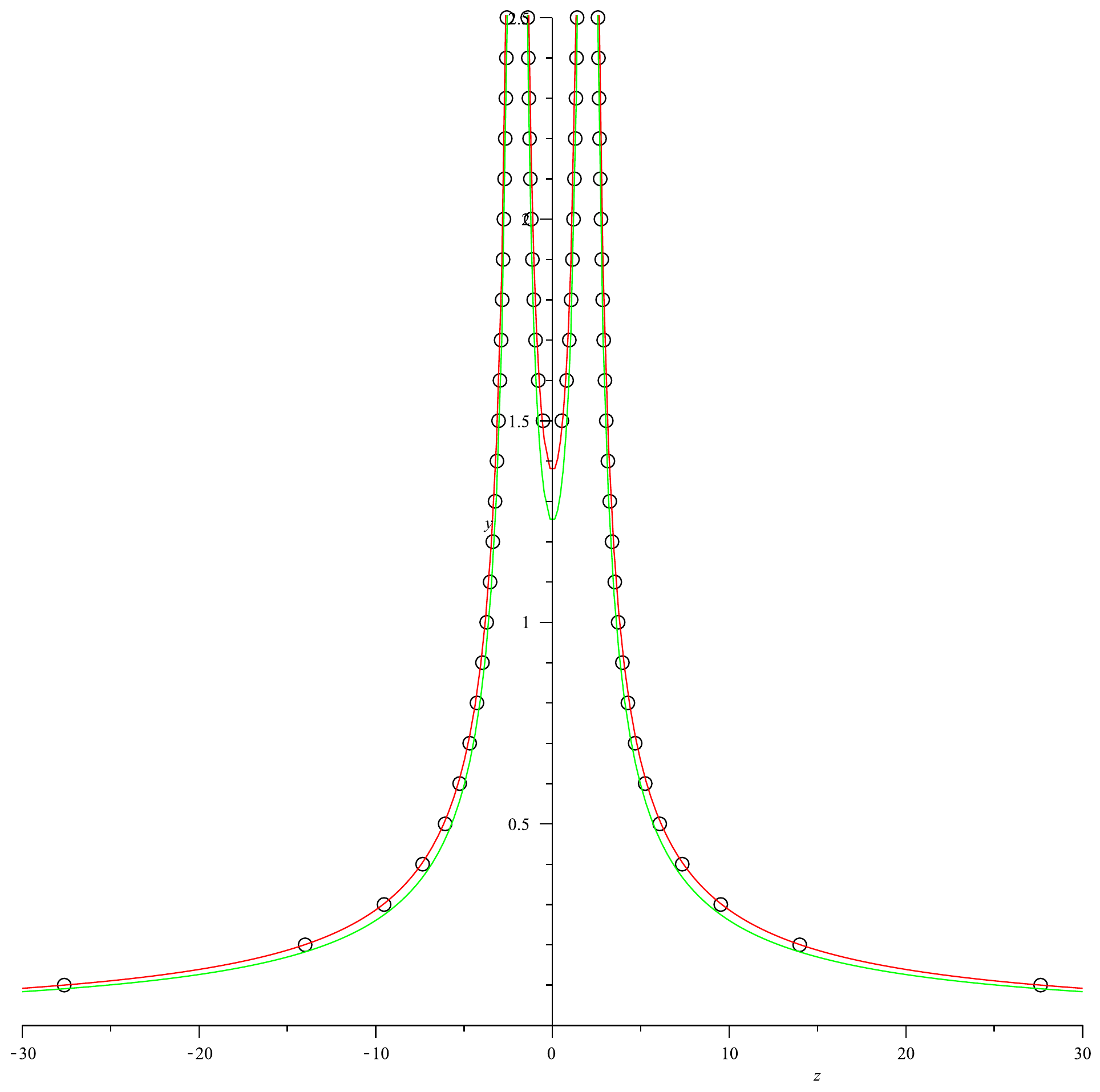}
\includegraphics[width=0.4\textwidth]{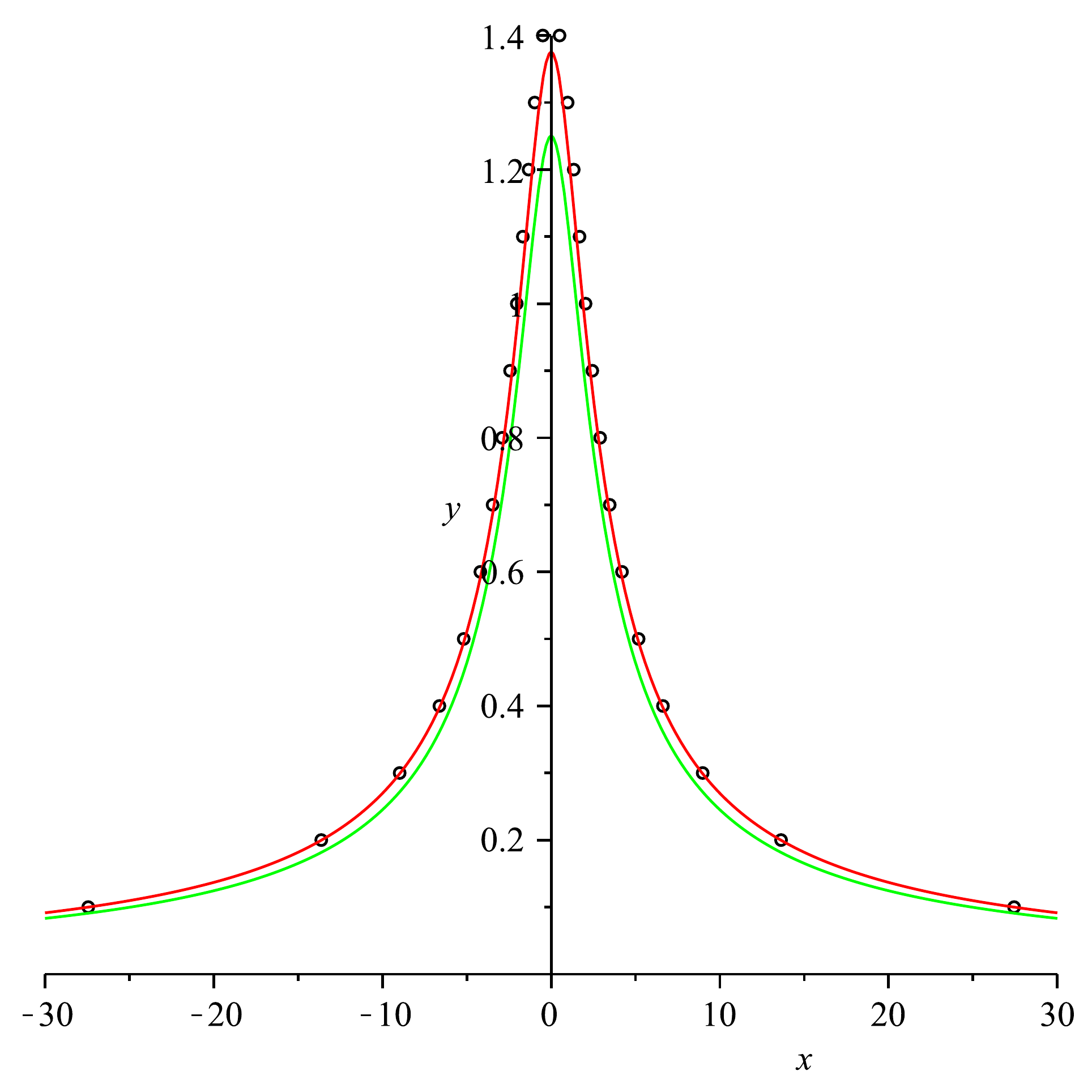}
\caption{$\sigma(x^1, x^2,  x^3)$  shown as a function of $x^3$ (on the left) along the line 
$x^1=x^2=0$ and  as function of $x^1$ along $x^2=x^3=0$.  
The D1-brane bundles are separated along $z$, as shown in 
equation (\ref{eqn:bions2}).  Solid lines show the D3-brane picture
(red for $q=N-1/2$, green for $q=N-1$), while open circles show the D1-brane picture.  $N=6$.}
\label{fig:profile}
\end{figure}

\begin{figure}
\centering
\includegraphics[width=0.8\textwidth]{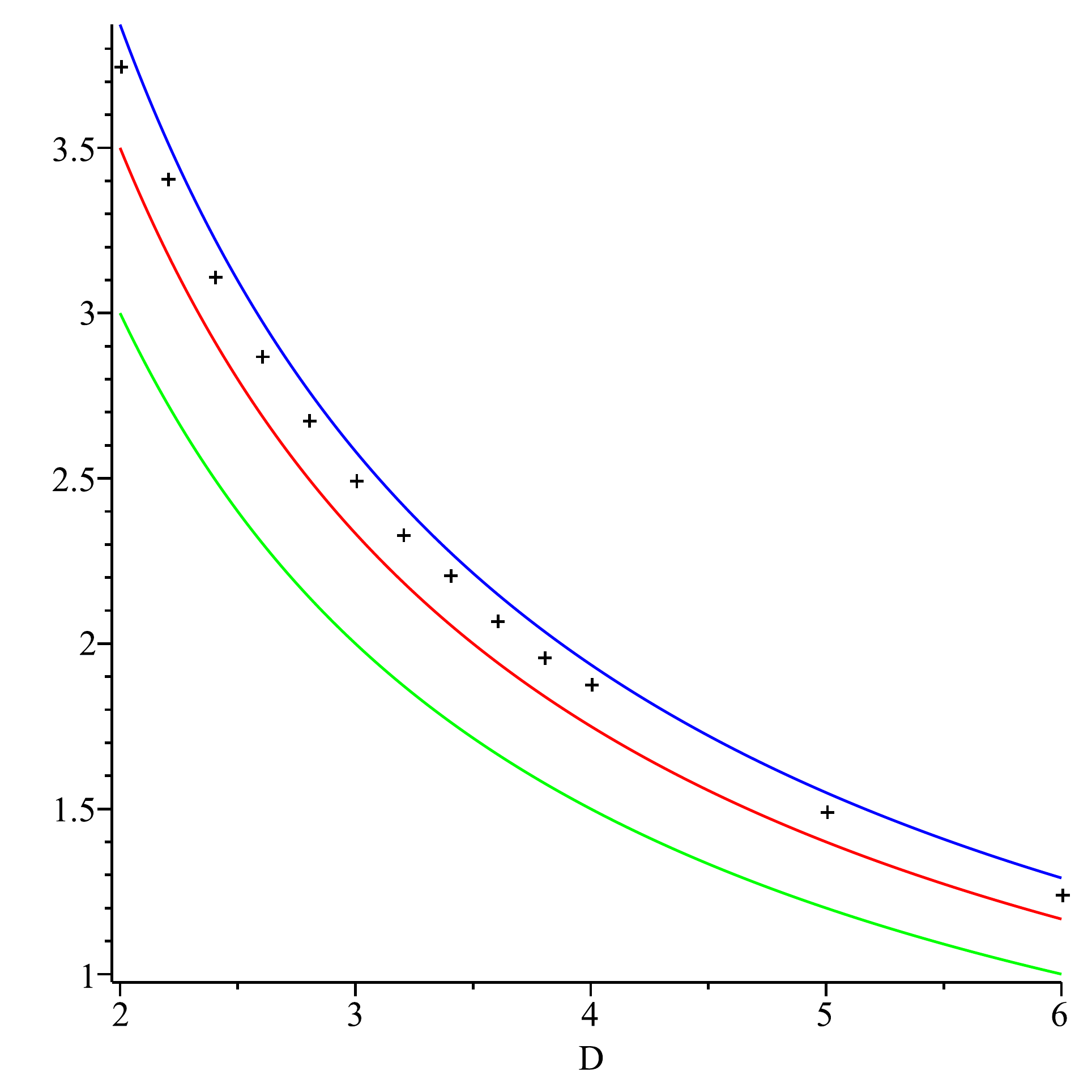}
\caption{$\s(0,0,0)$ as a function of $D$, separation of the D1-brane bundles.
Solid lines:  as predicted by the D3-brane picture (equation (\ref{eqn:bions2}),
blue $q=\sqrt{N^2-1}$, red $q=N-1/2$, green $q=N-1$);
points:  as obtained in the nonabelian geometry of the D1-branes. $N=4$.}
\label{fig:saddle-point}
\end{figure}

To examine the intermediate region in more detail, Figure \ref{fig:profile}
shows $\sigma(x^1, x^2,  x^3)$ along two lines: $x^1=x^2=0$ and $x^2=x^3=0$.
We see that the D1-branes reproduce the shape of the D3-brane 
with surprising accuracy, especially considering that $N$ is only $6$.  
Further, in Figure \ref{fig:saddle-point},
$\sigma(0,0,0)$ as a function of the separation of the D1-brane bundles is shown.
Again, reasonable agreement is seen, with the D1-brane picture falling somewhere between
the D3-brane predictions for $q=N-1/2$ and $q=\sqrt{N^2-1}$.  

We would expect the 
agreement between the D3-brane picture and D1-brane picture to get better as $N$ increases;
however, computations beyond $N=6$ are difficult, as evaluating the constants of motion
gets cumbersome for larger values of $N$.

\section{Future and related work}
\label{sec:last}

We have found novel solutions to the nonabelian Born-Infeld theory on a world-volume of
D1-branes.  These nonabelian configurations correspond to emergent geometry
exhibiting a topology change and are in excellent agreement with the dual D3-brane
picture.  The agreement we see between the D3-brane geometry and the emergent D1-brane
geometry can be viewed as a validation of the main result of work \cite{Berenstein:2012ts}.
It is perhaps surprising that the agreement should be this close given the relatively
small values of $N$ we have used.

Given these encouraging results, it would be worth while to consider some of the 
following extensions of our work:
\begin{itemize}
\item It would be very interesting to find a direct connection between the abelian Born-Infeld
action for D3-branes and the nonabelian Born-Infeld action for D1-branes
(or, similarly, the abelian Born-Infeld action for D2-branes and the nonabelian
Born-Infeld action for D0-branes) using the geometric interpretation of the 
nonabelian configurations.  This might be useful for resolving 
ordering difficulties in nonabelian Born-Infeld actions.
\item The moduli space of BPS solutions is very easy to understand from
the D3-brane picture, and quite obscure in the D1-brane picture.
Is there some feature of the Nahm equation 
which allows one to see the full moduli space of solutions?
\item Our solution technique could be extended to ever more
complex scenarios.  One could investigate several bundles of D1-branes, 
higher dimensional intersections (D1-D5 or D1-D7, for example, see
\cite{Cook:2003rx,Constable:2002yn,Constable:2001kv,Constable:2001ag}),
or, more ambitiously, multiple D3-branes and nonabelian the bion solutions, 
recently of interest.
\item Previous work on D-brane intersections includes
studies of fluctuations and thermal properties (see for example,
\cite{Bhattacharyya:2005cd,Grignani:2010xm,Grignani:2011mr}).  It could be interesting 
to extend this work to multiple bions.
\end{itemize}

Finally, we should mention previous work on related subjects.
The appearance of the Nahm equation as a BPS condition for D1-branes
is quite natural and provides an interpretation of the Nahm
procedure in terms of lower dimensional branes, as was first pointed
out in \cite{Diaconescu:1996rk}. The Nahm equation arises in many
contexts and is solved with different boundary conditions, dictated
by the problem at hand. The standard boundary conditions  are those
which are useful in the ADHMN construction of monopoles. There, $\s$
is taken on the interval $(-1,1)$ and the matrices $\Phi^i(s)$ have
poles at $\s = \pm 1$ whose residues are generators of the same
irreducible representation of $SU(2)$.  This corresponds to a bundle
of D1-branes connecting two parallel D3-branes separated by a finite
distance. By removing one of these poles to infinity so that $\s$
lives on the interval $(0,\infty)$, we remove one of the two branes
to infinity, and obtain the single D3-brane scenario described in
Section \ref{sec:nahm}. More complicated boundary conditions,
describing the discontinuity as a bundle of D1-branes crosses a
D3-brane were discussed in 
\cite{Tsimpis:1998zh,Kapustin:1998pb,Chen:2002vb}.

The problem of solving the Nahm equations with
different representations at $\s=0$ and $\s=\infty$
was considered in \cite{Bachas:2000dx}.
There the boundary conditions
(\ref{eqn:spikes}) and (\ref{eqn:brane})
with all $x^i_{(a)}=0$ were considered.
In the geometrical language of this paper, there was no
separation between the individual D1-brane bundles ($\Delta=0$).
Dimensions of the moduli space of solutions were computed
(for example, for $2 + 2 \rightarrow 4$ and $\Delta=0$ the
moduli space is 4-dimensional).

Separating D1-branes ending on the same D3-brane was also considered
in \cite{Chen:2002vb}, where the boundary condition for removing
one D1-brane from the bundle was considered.

%%%%%%%%%%%%%%%%%%%%%%%%%%%%%%%%%%%%%%%%%%%%%%%%%%%%%%%%%%%%%%%%%%%%
%  END MATTER: BIBLIOGRAPHY, ACKNOWLEDGMENTS, ...                  %
%%%%%%%%%%%%%%%%%%%%%%%%%%%%%%%%%%%%%%%%%%%%%%%%%%%%%%%%%%%%%%%%%%%%

\section*{Acknowledgments}
We would like to thank Cameron Funnel for his contributions to the early
stage of this project \cite{FunnelThesis}, and Gordon Semenoff for useful discussions.
This work was funded by the Natural Sciences and Engineering Resarch
Council of Canada.

\bibliographystyle{JHEP}
\bibliography{my}

\end{document}